# Classical perception follows solely from the linearity of quantum mechanics with no need for decoherence or the environment.

Casey Blood*
CaseyBlood@gmail.com

## Abstract

As illustrated by Schrödinger's cat, there are often several macroscopically different versions of reality simultaneously existing in the wave function. On the face of it, this would seem to imply that an observer could perceive a superposition of versions and thereby put quantum mechanics at odds with our perceptions. However one can show, without invoking decoherence or the environment, that the linearity of quantum mechanics implies we will perceive only a single, classical version of reality. Linearity also implies our perceptions will have a particle-like consistency.

## I. Introduction

Quantum mechanics is an astonishingly successful theory. It gives predictions, accurately verified by experiment, in elementary particles, atomic and nuclear physics, and the properties of solids, including the semiconductors used in all electronic devices. Further, in the many thousands of comparisons between theory and experiment, it has never given a wrong answer. In spite of these quantitative successes, however, the theory has what appears to be a severe shortcoming. Quantum mechanics is a mathematical theory in which objects are represented by or associated with wave functions/state vectors (whose nature we don't fully understand). The problem is that, because of the linearity of the theory, the state vector often contains two or more macroscopically different versions of reality. In the Schrödinger's cat experiment [1], for example, the cat is simultaneously dead and alive. That is, there is one version of reality in the state vector that properly describes what we would perceive if the cat were alive; and, at the same time, there is also a version of reality that properly describes what we would perceive if the cat were dead!

If we think of the observer as simply looking at the state vector, the simultaneous existence of two or more macroscopically different, equally valid versions of reality in the state vector would seem to imply the observer would perceive superpositions of classical versions of reality rather than just the single-version classical reality we actually perceive. That is, we should perceive some odd mixture of cat alive and cat dead. If this is actually what quantum mechanics implies, then there is a conflict between the theory and our everyday perceptions. This is often called the measurement problem [2] because the multiple realities in the wave function after a measurement do not match our perception of a single reality.

Over the many decades since quantum mechanics was discovered, physicists have attempted to resolve this conflict by suggesting various modifications of or additions to the mathematics. One of the major attempts is collapse of the wave function; it is assumed the wave function, for unknown reasons, collapses down to just one of the possible versions of reality [3-5]. There would then be an objective reality made up of the wave function corresponding to one of the versions. Experimental tests of this proposal have yielded no evidence for it, and have in fact put substantial restrictions on possible collapse theories [6-8].

The other major attempt to get around the measurement problem is to suppose the wave function is not the actual perceived reality. Instead, there is presumed to be an objective single-version reality made up, perhaps, of real protons and electrons, with the state vector's role being to *describe* the reality rather than to *be* the reality. There have been several attempts to formulate a theory of the physical world based on this idea, often going under the name of "hidden variables" [9-11]. But when experimentally tested, none of the attempts—including Bell-like experiments [12,13]—has succeeded in showing there is an objective reality. And as in the collapse case, they have established restrictions on possible hidden variable theories.

But neither of these attempts to solve the measurement problem is necessary because when you look closely, it turns out that quantum mechanics does not lead to a contradiction with our perceptions after all. In spite of the state vector containing several versions of reality, the mathematics never allows the *perception* of anything other than a single version of reality. And that single perceived version is "classical" in the sense that the cat will be perceived either as dead or as alive, but will never be perceived as a mixture or superposition of the two.

The strategy of resolving the apparent conflict by showing quantum mechanics itself implies classical perception, has been proposed by Zeh [14,15] using decoherence and particularly by Zurek [16-21] who uses the information contained in the environment. But their methods miss the essential reason for seeing only classical results, for as we will show, classical perception is implied solely by the linear time evolution of the state vector, with no reference to decoherence or the environment. In addition neither Zeh nor Zurek include the state of the observer and that makes it more difficult to evaluate their arguments.

Our approach is similar to Everett's relative state or "many-worlds" formulation [22] of quantum mechanics because we are considering the perceptual consequences of presuming only the wave functions/state vectors exist. But Everett does not fully come to grips with the measurement problem.

The effects of a measurement on a system are discussed in Sec. II. In Sec. III, classical perception is defined and the role of the observer is discussed. The proof of the primary result—that no version of the observer ever perceives a non-classical result—is given in Sec. IV. It uses only the linear time evolution of the system to show that the *perception* of Schrödinger's cat as a mixture of alive and dead cannot occur in quantum mechanics. In Sec. V, again using only linearity, it is shown that observers agree on their perceptions and that consecutive measurements give consistent results. In addition, spread-out wave functions are perceived as being localized to a small region and as giving particle-like trajectories in a bubble chamber. Finally we summarize in Sec. VI and note that these results invalidate several of the arguments used to justify the alleged collapse of the wave function or the existence of particles or hidden variables.

## II. Post-measurement states

We need to establish the general form of the particle-detector state vector after detection (where "particle" is shorthand in this paper for "particle-like wave function"). Suppose we consider a scattering experiment in which the compound detector is a set of N film grains coating a sphere surrounding the scattering center, with initial state vector

$$|\Psi_0\rangle = |\mathrm{pa}\rangle \prod_{i=1}^{N} |gr(i), n\rangle \quad (1)$$

where $|gr(i), n\rangle$ stands for the not-exposed grain *i*. The time translation operator is

$$U(t) = \prod_{j=1}^{N} \exp(itH_j) = \prod_{j=1}^{N} (B(j)+1), \; B(j) = \exp(itH_j) - 1 \quad (2)$$

where $H_j$ is the short-range potential energy for the interaction between the particle and the $j^{\text{th}}$ grain. If the particle wave function is spread out in space, then $H_j$ acting on $|pa\rangle|\mathrm{gr}(\mathrm{j})\rangle$ is 0 except for the part of the particle wave function, $|pa, x_j\rangle$, in the immediate vicinity of grain *j*. Thus

$$B(j)B(k)|\Psi_0\rangle = 0, \; k \neq j \, . \quad (3)$$

because *B*(*k*) localizes the particle wave function to a region where $H_j$, and thus *B*( *j*), are 0. (More precisely the result in (4) follows when (3) holds to high accuracy, which it does in virtually all situations of interest.) So if we expand the product in (2), taking into account that all cross terms give 0 and that the particle-grain interaction $H_j$ exposes grain *j* and only grain *j*, we have

$$|\Psi_f\rangle = \mathrm{U}(\mathrm{t})|\Psi_0\rangle = \sum_{j=1}^{N} b_j \, |\mathrm{pa}, x_j\rangle \{|\mathrm{gr}(\mathrm{j}), y\rangle \prod_{k \neq j} |gr(k), n\rangle\} \quad (4)$$

where the *y* stands for yes, exposed, the $b_j$ are normalization constants (so the particle state is normalized to 1) and the states in curly brackets are the N *post-measurement states* of the detectors. The final state is now a sum of terms or branches in each of which one and only one grain is exposed, and the wave function for the particle states is non-zero only in the region of the exposed grain. Thus the position of the particle after detection is correlated with the readings on the detectors.

If instead we had considered a Stern-Gerlach-like experiment with initial state $|\Psi_0\rangle = \sum_i a_i \, |pa, i\rangle \prod_j |\det j, n\rangle$ where $|\det j, n\rangle$ stands for the untriggered detector *j*, then the final state is similar to (4);

$$|\Psi_f\rangle = \sum_i a_i \,|\,\text{pa}, x_i\rangle \{|\det i, y\rangle \prod_{j\neq i} |\det j, n\rangle\}\,. \qquad (5)$$

The detector readings in the N different post-measurement states each correspond to a classical outcome with one and only one grain exposed, and the particles are localized to a region near the "yes" detector. Equations (4) and (5), with their entanglement of the particle and detector states, constitute a significant part of the journey from quantum mechanics to classical, particle-like observations.

## III. Classical perception. The observer

We consider a quantum mechanical experiment, with macroscopic detectors, in which there are a finite number of mutually orthogonal post-measurement states of the detectors, as in (4) and (5). A classical perception, by definition, is the perception of the detector readings corresponding to just one of those states. In the Schrödinger's cat example, the two classical perceptions are of course cat alive and cat dead. And in a double slit or scattering experiment with N film grains as the detectors, the N classical perceptions each correspond to the perception of just one exposed grain.

Now the observer. Our goal is to show that pure, no-collapse, no-hidden variable, multi-version quantum mechanics leads to classical, single-version *perception* by the *observer*. So the observer and her perceptions must be included in our deliberations. (One might also argue that the state of the observer must be included because the observer is not just *looking at* the quantum state; the observer is also *part of* the quantum state.) The observer is presumed to be entirely described by quantum mechanics so it is simply an inanimate device which "perceives" the outputs of the detectors and responds in a way to be described in a moment.

### A. The logic

After an experiment is run and the results observed, there will be several different versions of the observer; there is no singular *the* observer in no-collapse, no-hidden variable quantum mechanics. What we will show— without the assumption of particles or collapse, or the use of decoherence or the environment—is that *no version of the observer ever perceives a non-classical result*. This is sufficient to show that the linearity of quantum mechanics, by itself, leads to our classical everyday perceptions. It shows that you, as the observer, will perceive Schrödinger's cat as either dead or alive, but you will never perceive some combination of the two. (It doesn't, however, predict which classical version you will perceive.)

The reasoning might also be stated a little differently. If one is to claim that no-collapse, no-particle quantum mechanics leads to non-classical perceptions, then it is reasonable to require that a specific example be given in which the time evolution of the system leads to non-classical perception by some version of the observer. But the argument of Sec. IV will show it is impossible to construct such an example.

## IV. Derivation of classical perception

We start the derivation from the state of (4) with an observer added to the system. At time $t_1$, after the quantum mechanical system has been detected but before the observer perceives the results, the state vector is

$$|\Psi(t_1)\rangle = \sum_{i=1}^{N} b_i\ |i?, t_1\rangle\ |\mathcal{D}(i,t_1)\rangle\ |\mathrm{Obs},0\rangle \quad \sum_{i=1}^{N} |b_i|^2 = 1. \tag{6}$$

The possible states of the particle at $t_1$ are $|\mathrm{i}?, t_1\rangle$ with the "?" signifying that this term is not there if we are dealing with photons that are annihilated. $|\mathrm{Obs},0\rangle$ represents the state of the observer before perceiving the readings on the detectors. And $|\mathcal{D}(i,t_1)\rangle$ represents the $i^{\text{th}}$ state of the detectors after detection so we are, in effect, lumping all the detectors into just one "super-detector." If we do a double slit or scattering experiment for example, the detectors might consist of N film grains, with the detector state $|\mathcal{D}(i,t_1)\rangle$ corresponding to grain *i* being exposed and the rest of the grains unexposed, as in (4).

After observation, the observer is to select one of N+1 choices; choice *i* is selected if she perceives classical outcome *i*, and choice N-C is selected if she perceives a non-classical result. (It is of course not necessary for the observer to be a "conscious" entity in order to make such a choice.) The state after the observer looks, at time $t_2$, is

$$|\Psi(t_2) = \mathrm{U}(t_2,t_1)\ |\Psi(t_1)\rangle = \tag{7}$$

$$\sum_{i=1}^{N} b_i\ \mathrm{U}(t_2,t_1)\ \{|\mathrm{i}?,t_1\rangle\ |\mathcal{D}(i,\ t_1)\rangle\ |\mathrm{Obs},0\rangle\}$$

where U is the time evolution operator. Now the action of $\mathrm{U}(t_2,t_1)$ on $\{|\mathrm{i}?,t_1\rangle\ |\mathcal{D}(i,\ t_1)\rangle\ |\mathrm{Obs},0\rangle\}$ is independent of the coefficients so we can choose $b_i$=1, $b_j$=0, $j \neq i$ in evaluating this term. But since the *j* states are no longer there in such a choice, the version of the observer on the $i^{\text{th}}$ branch can only perceive state *i*, with perception being mediated by the photons that travel from the detectors in state *i* to the observer's eyes. Thus

$$|\Psi(t_2)\rangle = \sum_{i=1}^{N} b_i\ |\mathrm{i}?, t_2\rangle\ |\mathcal{D}(i,t_2)\rangle\ |\mathrm{Obs\ selects\ switch}\ i\rangle. \tag{8}$$

We see that N-C is never selected so *no version of the observer ever perceives a non-classical state*! The time evolution from the state of (6) to perception guarantees that versions of the observer perceive only classical outcomes. Thus the multi-version state vector leads to single-version perceptions, in agreement with our daily observations.

### A. Change of basis

Is it possible to circumvent this reasoning by changing bases? We will show by considering a two-state system, with a detector that reads 1 or 2, that it is not. The state after perception is

$$|\Psi\rangle = a_1|\mathrm{pa},1\rangle\ |\mathrm{Det},1\rangle\ |\mathrm{Obs\ selects}\ 1\rangle + a_2|\mathrm{pa},2\rangle\ |\mathrm{Det},2\rangle\ |\mathrm{Obs\ selects}\ 2\rangle \tag{9}$$

$$= a_1\ |\mathrm{D},1\rangle\ |\mathrm{O},1\rangle + a_2\ |\mathrm{D},2\rangle\ |\mathrm{O},2\rangle.$$

The new basis vectors for the observer are

$$|O, 1'\rangle = \cos(\theta)\,|O,1\rangle - \sin(\theta)\,|O,2\rangle \qquad (10)$$

$$|O, 2'\rangle = \sin(\theta)\,|O,1\rangle + \cos(\theta)\,|O,2\rangle$$

or

$$|O, 1\rangle = \cos(\theta)\,|O, 1'\rangle + \sin(\theta)\,|O, 2'\rangle \qquad (11)$$

$$|O, 2\rangle = -\sin(\theta)\,|O, 1'\rangle + \cos(\theta)\,|O, 2'\rangle .$$

Substituting (11) into (9) gives

$$|\Psi\rangle = \{a_1\,|D,1\rangle\cos(\theta) - a_2\,|D,2\rangle\sin(\theta)\}\;|O, 1'\rangle + \qquad (12)$$

$$\{a_1\,|D,1\rangle\sin(\theta) + a_2\,|D, 2\rangle\cos(\theta)\}\;|O, 2'\rangle .$$

Thus the $|O,1'\rangle$ and $|O,2'\rangle$ versions of the observer *appear* to perceive non-classical versions of reality. But *there is no reason to equate the product form*,

$$\{a_1|D, 1\rangle\cos(\theta) - a_2|D, 2\rangle\sin(\theta)\}\;|O, 1'\rangle , \qquad (13)$$

with *perception* of the state in curly brackets by the $|O,1'\rangle$ version of the observer. In fact, we see from (10),

$$|O,1'\rangle = \cos(\theta)|\text{Obs selects}1\rangle - \sin(\theta)\,|\text{Obs selects}2\rangle \qquad (14)$$

$$|O,2'\rangle = \sin(\theta)|\text{Obs selects}1\rangle + \cos(\theta)\,|\text{Obs selects}2\rangle ,$$

that N-C *is still not selected in this basis*. The conclusion is that, in spite of the validity of all bases in quantum mechanics and the fact that there are several versions of reality in the wave function, no quantum mechanical version of the observer will ever perceive a non-classical result. Thus quantum mechanics guarantees the world will be perceived as classical.

One might ask: If (12) does not represent perception of "mixed" states by the versions $|O,1'\rangle$ and $|O,2'\rangle$ of the observer, what *does* it represent? The answer is that, because the left hand sides of (9) and (12) represent the same physical state, the two right hand sides must also represent the same state (even though they are expressed in different ways). And that state, for both (9) and (12), corresponds to perception of detector state $|\text{Det } 1\rangle$ by observer version $|O, 1\rangle$ and perception of detector state $|\text{Det}, 2\rangle$ by version $|O, 2\rangle$.

In an attempt to show non-classical perception, we could also try changing the basis for the particle-detector states. But because of the linearity of $U(t_2,t_1)$, this again does not lead to non-classical perception.

To summarize each of us, as an observer, is a quantum object that never signals N-C and thus can never be said to perceive a non-classical result.

## V. Classical particle-like consistency of perception

We have shown that no version of the observer ever perceives anything other than a single, classical version of reality which corresponds to one of the post-measurement states. But there are other classical properties which perception in quantum mechanics should have if it is to fully imitate a classical world. We will describe these here and show they do indeed follow from linearity.

### A. Agreement among observers

The first is agreement among observers. To illustrate, suppose we do a Stern-Gerlach experiment with two observers. The particle (particle-like wave function) can travel on either path 1 or path 2, with detectors D1 on path 1 and D2 on path 2. Observer A has two switches labeled 1, 2, and observer B has four switches 1, 2, agree, disagree.
The state vector at time 0, after detection but before perception is

$$\begin{aligned}
&\Psi(0)= a_1\Phi_1(0)+a_2\Phi_2(0)\\
&\Phi_1(0)=|\text{pa on path 1}\rangle|\text{D1}(y)\rangle\,|\text{D2}(n)\rangle\,|\text{Obs A},0\rangle\,|\text{Obs B},0\rangle\\
&\Phi_2(0)=|\text{pa on path 2}\rangle|\text{D1}(n)\rangle\,|\text{D2}(y)\rangle\,|\text{Obs A},0\rangle\,|\text{Obs B},0\rangle
\end{aligned} \tag{15}$$

And the state vector at time t is

$$\Psi(t)= a_1\,U(t)\,\Phi_1(t) + a_2\,U(t)\,\Phi_2(0) = a_1\,\Phi_1(t)+ a_2\,\Phi_2(t) \tag{16}$$

But the two $\Phi$ s are effectively in separate universes (in Hilbert space) and so they will evolve independently. Thus we must have

$$\begin{aligned}
&\Phi_1(t) = |\text{pa on path 1}\rangle\,|\text{D1}(y)\rangle\,|\text{D2}(n)\rangle\,|\text{Obs A sees } y,n\rangle\,|\text{Obs B sees } y,n\rangle \\
&\Phi_2(t) = |\text{pa on path 2}\rangle\,|\text{D1}(n)\rangle\,|\text{D2}(y)\rangle\,|\text{Obs A sees } n,y\rangle\,|\text{Obs B sees } n,y\rangle
\end{aligned} \tag{17}$$

Thus because versions of the observer cannot perceive "across branches," the versions of observer B on both branches will choose "Agree." This implies there will never be disagreement among observers on the perceived branch.

### B. Perceived locality of particle-like wave functions

Particles in a classical world are typically assumed to be localized in a (single) small region of space. To show that perception in (no-particle) quantum mechanics imitates this property, suppose we perform a scattering or double slit experiment in which the N detectors are N film grains. The argument of Sec. IV shows that one and only one of the post-measurement states of (4), $|gr(j),y\rangle\prod_{k\neq j}|gr(k),n\rangle$, will be perceived. So we see that even though the wave function is spread out and hits all the grains, the *perceived* result is *just as if* a small, classical particle hit and exposed a single grain!

### C. Particle-like trajectories. Classical consistency

There are situations where successive measurements are made on a wave function. The most common would be the particle-like trajectories observed in a bubble chamber, where each potential nucleation site is effectively a detector. We will show that the perceived outcomes in such a situation are just what one would expect if there really were particles.

We suppose there are two sets of detectors, DA($i$) with $i$ running from 1 to $N_A$ and DB($j$), with $j$ from 1 to $N_B$. If we had a scattering experiment, for example, the DA($i$) would correspond to film grains on one sphere and the DB($j$) to film grains on a slightly larger sphere. After the wave function hits and passes the inner sphere, but before it reaches the outer sphere, the state vector is (see (4))

$$[\sum_{i=1}^{N_A} a(i)\,|\,\mathrm{DA}(i,y)\rangle \prod_{i'\neq \mathrm{i}}^{N_A} |\mathrm{DA}(i',n)\rangle\,|pa,\,x_i\rangle]\ [\prod_{j=1}^{N_b} |\mathrm{DB}(j)\rangle] \tag{18}$$

Now when $|\mathrm{pa},\,x_i\rangle$ hits sphere B, it can only activate the B detectors which are in the path of $|pa,\,x_i\rangle$, and the particle wave function then becomes localized near the B detector it activates. So the full state vector, after the wave function hits both detectors, is

$$[\sum_{i=1}^{N_A} a(i)\ ||\mathrm{DA}(i,y)\rangle \prod_{i'\neq \mathrm{i}}^{N_A} |\mathrm{DA}(i',n)\rangle\,] \tag{19}$$

$$[\sum_{j=1}^{N_B} a'(i,j)\ |\mathrm{DB}(j,y)\rangle \prod_{j'\neq \mathrm{j}}^{N_B} |\mathrm{DB}(j',n)\rangle]\ |\mathrm{pa},\,x_i\,,\,x_j\rangle]$$

where the "trajectory" of the state vector $|\mathrm{pa},\,x_i\,,\,x_j\rangle$ first went through detector DA($i$) and then through detector DB($j$) and ends up (for this two-step analysis) near DB($j$). Thus for each perceivable outcome, there is one activated detector in the A sphere, and one in the B sphere, just as we classically expect. Further, if the A detector does not significantly alter the trajectory of the particle-like state, then most of the $a'(i,j)$ are zero; only those $j$ grains with detector DB($j$) nearly behind DA($i$) have the potential to be activated. Thus, because of (4), the linear mathematics implies the perceived states of the detectors will be *just as if* a localized particle hit DA($i$) and then hit DB($j$) behind it. An extension of this to many spheres of detectors instead of just two shows that spread-out particle-like wave functions going through a bubble chamber leads to the perception of a classical (that is, sharply defined, not disjoint), particle-like trajectory [23].

There is one other situation where this classical behavior is of particular interest and that is repeatability, which is one of the postulates that is sometimes *postulated* [18,19,24] in trying to match quantum mechanics with classical perception. To explain, suppose we do the Stern-Gerlach experiment and perceive results corresponding to the $|+\rangle$ state. Then to show repeatability one must show that quantum mechanics also leads to perception of the $|+\rangle$ outcome if a second, successive measurement is made. Classically, this is equivalent to saying that if a particle is measured to be in a certain state, then a second measurement should confirm that.

But there is no need to postulate this property because it follows from linearity. (Actually, this is just a special case of (19) but we will consider it more explicitly.) To show this, we do the spin ½ Stern-Gerlach experiment with two detectors, DA(1), DA(2) and DB(1), DB(2), on paths A and B resp. with the "2" detectors positioned after the "1" detectors. The observer has the choice of two switches, labeled "Agree" if DA(2), DB(2) are perceived as agreeing with DA(1), DB(1), and "Disagree" if they don't. The $|pa, x_i\rangle$ of (5) implies the part of the wave function that is detected by DA(1) will continue on path A and activate DA(2); and the same for path B. So the final state implied by the linear time evolution is

$$\begin{aligned} &a|DA(1,y)\rangle|DA(2,y)\rangle|DB(1,n)\rangle|DB(2,n)\rangle|+\rangle_{\text{path A}}|\text{Obs selects Agree}\rangle+ \\ &b|DA(1,n)\rangle|DA(2,n)\rangle|DB(1,y)\rangle|DB(2,y)\rangle|-\rangle_{\text{path B}}|\text{Obs selects Agree}\rangle \end{aligned} \tag{20}$$

Disagree—meaning the perception of non-matching results for the A and B detectors—is never selected and so we have shown that repeatability follows from the basic rules of quantum mechanics. Zurek[16-21] takes repeatability as a postulate in his derivation of classicality from quantum mechanics, but we see it is actually a property that can be derived from the linearity of quantum mechanics.

## VI. Summary

The state vector for an experiment with N possible outcomes is a linear combination of states corresponding to those N possibilities. But the linearity of quantum mechanics implies the outcomes *perceived* by the quantum versions of the observer always correspond to just one of the N *classical* outcomes. No version of the observer ever perceives a non-classical result in which several classical outcomes are superimposed. Thus the observer sees Schrödinger's cat either as alive or as dead, but never as some combination of alive and dead. Further, linear, no-collapse, no-particle quantum mechanics leads to agreement among observers, particle-like consistency of perception for consecutive measurements, the perceived locality of spread-out wave functions, and the perception of classical, particle-like trajectories. No use is made of decoherence or the environment in obtaining these results. It is remarkable that perception in multi-version quantum mechanics, with its spread-out wave functions, so closely imitates perception of a single-reality classical world made up of highly localized particles.

Finally we note that, because these five phenomena—classicality of perception, agreement among observers, perceived particle-like locality of wave functions, particle-like trajectories, and consistency of consecutive results,—all follow from no-particle, no-collapse quantum mechanics, none of them may be used as evidence for collapse of the wave function or the existence of particles or hidden variables. If one does assume either collapse or hidden variables, then there are *two* explanations for each of these five phenomena. Nature is usually more parsimonious than that.

## References

*Present address. 708 S. American Street, Philadelphia, PA 19147.

[1] E. Schrödinger, Naturwissenschaften, **23**: 807 (1935).

[2] Eugene P. Wigner, Am. J. Physics, **31**, 6, (1963).

[3] W. Heisenberg. Z. Phys. **43**: 172 (1927).

[4] J. von Neumann. *Mathematische Grundlagen der Quantenmechanik* (in German). (Springer, Berlin 1932)

[5] G. C. Ghirardi, A. Rimini, and T. Weber, Phys. Rev. D **34**, 470 (1986); Phys. Rev. D, **36**, 3287 (1987).

[6] Philip Pearle, arXiv, quant-ph/0611211v1 (2006).

[7] Philip Pearle, arXiv, quant-ph/0611212v3 (2007).

[8] A. J. Leggett, J. Phys: Condens Matter **14** (2002).

[9] L. de Broglie, J de phys Radium" **8** (5): 225 (1927).

[10] David Bohm, Phys. Rev. **85** 166,180 (1952).

[11] D. Bohm and B. J. Hiley, *The Undivided Universe* (Routledge, New York, 1993).

[12] J. S. Bell, Physics, **1**, 195 (1964).

[13] A. Aspect, P. Grangier, and G. Rogers, Phys. Rev. Lett. **47**, 460 (1981).

[14] D. Giulini, E. Joos, C. Kiefer, J. Kupsch, I.-O. Stamatescu, and H.D. Zeh, in *Quantum Theory* (Springer, Berlin 1996).

[15] H. Dieter Zeh, "Decoherence: Basic Concepts and Their Interpretation," *arXiv:*quant-ph/9506020v3 (2002).

[16] W. H. Zurek, Phys. Rev. A **71**, 052105 (2005).

[17] Wojciech Hubert Zurek, arXiv:quant-ph/0405161v2 (2005).

[18] Wojciech Hubert Zurek, arXiv:quant-ph/0707.2832 (2007).

[19] Wojciech Hubert Zurek, Nature Physics, **5**, 181 (2009).

[20] C. Jess Riedel, Wojciech H. Zurek, Michael Zwolak, arXiv: quant-ph/1205.3197 (2012).

[21] Wojciech Zurek, Physics Today, **67**, 44, (2014).

[22] Hugh Everett, III, Rev. Mod. Phys. **29** 454 (1957). We note that Everett's state-vector-only interpretation cannot be the full story because it cannot properly account for probability.

[23] N. F. Mott, P. R. Soc. (London) **126**, 79 (1929).

[24] P. A. M. Dirac, *The Principles of Quantum Mechanics* 4th Edition, (Clarendon Press, Oxford, 1976).